\documentclass[%
 reprint,
 amsmath,amssymb,
 aps,
]{revtex4-2}

\usepackage{graphicx}
\usepackage{dcolumn}
\usepackage{bm}
\usepackage{float}
\usepackage{xcolor}

\begin{document}

\preprint{APS/123-QED}

\title{Valley protected one-dimensional states in small-angle twisted bilayer graphene}

\author{J.D. Verbakel\textsuperscript{1}}
\author{Q. Yao\textsuperscript{2}}%
\author{K. Sotthewes\textsuperscript{1}}
\author{H.J.W. Zandvliet\textsuperscript{1}}
\affiliation{%
\textsuperscript{1}Physics of Interfaces and Nanomaterials, MESA\textsuperscript{+} Institute for Nanotechnology, University of Twente, P.O. Box 217, 7500AE, Enschede, The Netherlands \\
\textsuperscript{2}Center for Artificial Low Dimensional Electronic Systems, Institute for Basic Science (IBS), Pohang 37673, Republic of Korea
}%


\begin{abstract}
Theory predicts that the application of an electric field breaks the inversion symmetry of AB and BA stacked domains in twisted bilayer graphene, resulting in the formation of a triangular network of one-dimensional valley-protected helical states. This two-dimensional network of one-dimensional states has been observed in several studies, but direct experimental evidence that the electronic transport in these one-dimensional states is valley-protected is still lacking. In this study, we report the existence of the network in small-angle twisted bilayer graphene at room temperature. Moreover, by analyzing Fourier transforms of atomically resolved scanning tunnelling microscopy images of minimally twisted bilayer graphene, we provide convincing experimental evidence that the electronic transport in the counter-propagating one-dimensional states is indeed valley protected. 
\end{abstract}

\maketitle

\begin{figure}[b]
	\centering
	\includegraphics[width=0.5\textwidth]{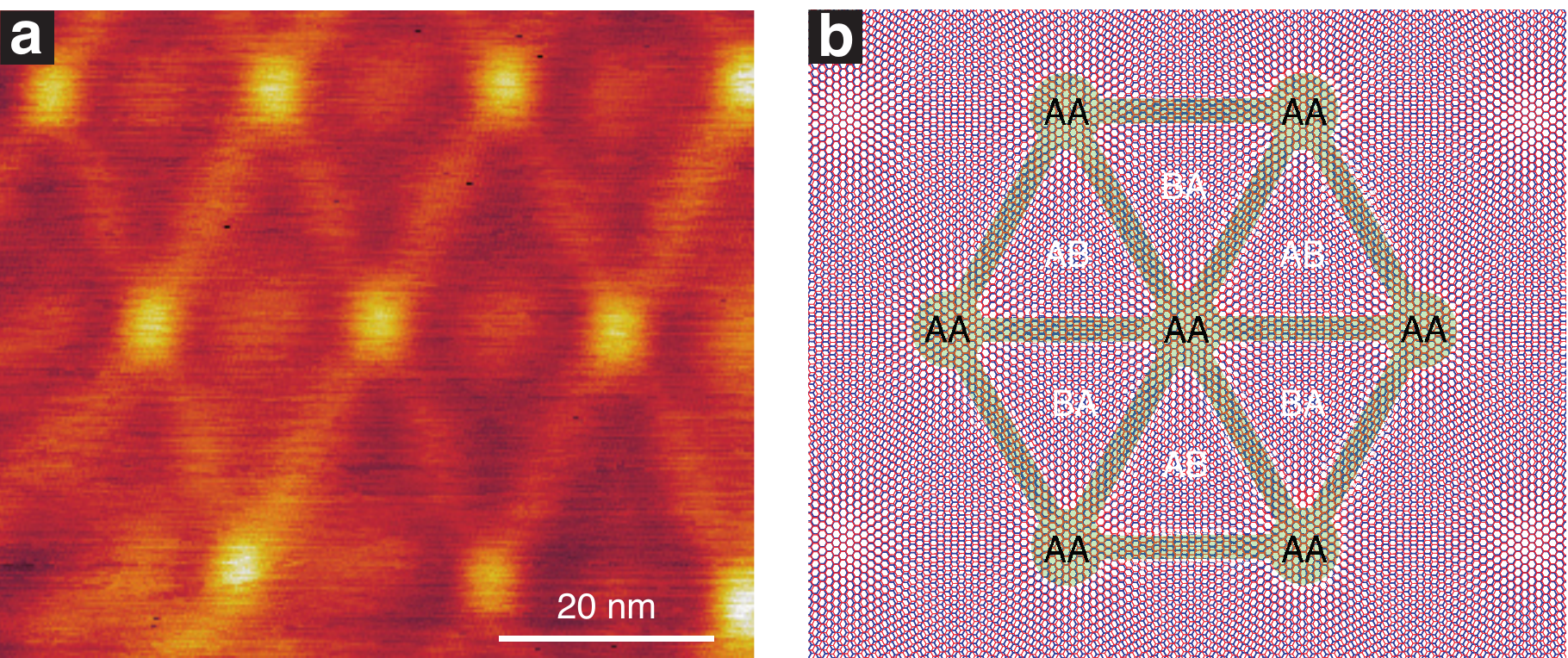}
	\caption{Network of topologically protected states in twisted bilayer graphene. 
		(a) STM image of the network of topologically protected states. The bias voltage is 400 mV and the current setpoint is 500 pA.
		(b) Schematic representation of the moiré pattern of twisted bilayer graphene, indicating the AA, AB and BA stacked regions. The green highlights indicate the triangular network.
	}
	\label{fig:network}
\end{figure}

The successful isolation and exploration of graphene~\cite{Novoselov2004,Geim2007} has led to a booming new research field in condensed matter physics. Graphene consists of a single layer of \textit{sp}\textsuperscript{2} hybridized carbon atoms, which are arranged in a honeycomb structure. Tight binding calculations of free-standing graphene \cite{Wallace1947,Neto2009} have revealed that the dispersion relation of the low-energy electrons bands is linear. The latter implies that the low-energy electrons in graphene behave as massless relativistic particles and are described by the Dirac equation, i.e. the relativistic variant of the Schrödinger equation. \\
\indent Bilayer graphene has a more complex band structure, which depends on the stacking order of the graphene layers. The most common and energetically most favorable stacking order is the so-called AB or Bernal stacking.  In the AB stacking configuration, the atoms of one of the triangular sub-lattices of the top layer (A\textsubscript{1}) are located on top of the atoms of one of the sub-lattices of the bottom layer (B\textsubscript{2}). The other atoms (B\textsubscript{1} and A\textsubscript{2}) do not lie directly below or above an atom of the other layer. For Bernal stacked bilayer graphene, the low-energy bands are not linear, but quadratic [5]. The electronic spectrum of bilayer graphene can be altered even more drastically by rotating one of the graphene layers with respect to the other. This 'twisting' of the bilayer graphene provides an additional degree of freedom, which allows the structural and electronic properties of the bilayer to be tailored. Twisting bilayer graphene results in a new superstructure, which is referred to as a moiré structure, in which domains of AA, AB and BA stacking exist, forming a triangular superlattice. The moiré pattern has a periodicity of $\lambda=a/[2\sin(\theta⁄2)]$, where $\theta$ is the twist angle and $a$ the lattice constant of graphene. \\
\indent Due to the hybridization of the Dirac cones of the top and bottom layer, the electronic band structure of twisted bilayer graphene (TBG) features two van Hove singularities, one located above and one located below the Fermi level~\cite{Li2010,Yan2012,Brihuega2012,VanHove1953}. By adjusting the twist angle, these van Hove singularities can be brought arbitrary close to the Fermi level. The latter allows one to systematically study and tune the formation of electronic instabilities and correlated electron phases, such as Mott metal-insulator transitions, Wigner crystallization and magnetism \cite{Cao2018,Yankowitz2019,Padhi2018,Sharpe2019,Gonzalez-Arraga2017}. At a twist angle of 1.1\textdegree, the so-called magic angle, the energy bands near the Fermi level become completely flat, resulting in a vanishing Fermi velocity. In 2018, Cao \textit{et al.} \cite{Cao2018} showed that at this magic twist angle the bilayer graphene exhibits superconductivity. 
At twist angles smaller than the magic angle, another interesting phenomenon has been reported: upon application of a perpendicular electric field, small-angle twisted bilayer graphene hosts a network of topologically protected states across the moiré pattern \cite{Huang2018,Rickhaus2018,Xu2019}. This network has previously been observed using scanning tunnelling microscopy (STM) by Huang \textit{et al.}~\cite{Huang2018} at temperatures of 4.5 K. However, our experiments show that this network can also be observed at non-cryogenic conditions: in figure~\ref{fig:network}, a room temperature STM measurement of the network in twisted bilayer graphene is shown. The twist angle of the bilayer is 0.55\textdegree, as extracted from the periodicity of the moiré pattern.
\begin{figure*}
	\centering
	\includegraphics[width=1\textwidth]{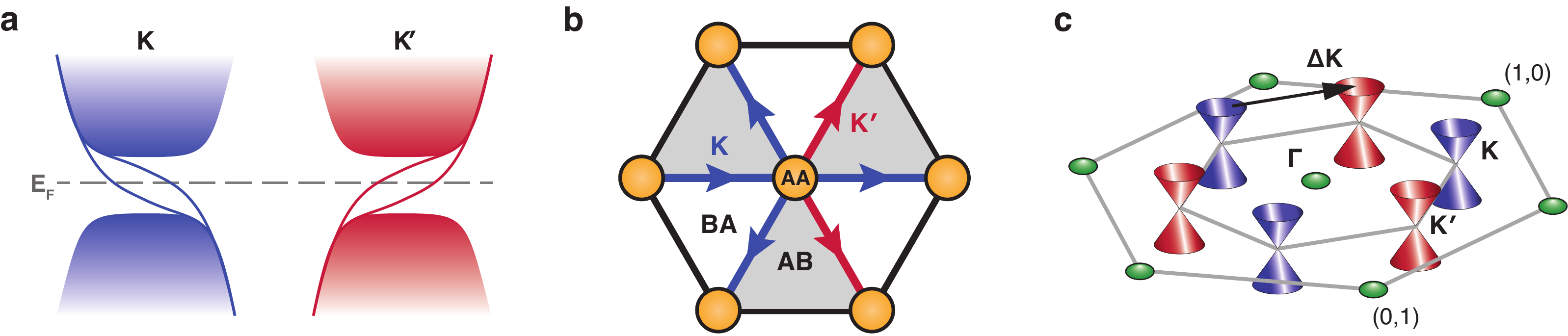}
	\caption{Network of topologically protected states in twisted bilayer graphene. 
		(a) Schematic of the band structure of twisted bilayer graphene under an external electric field in the $K$ and $K'$ valleys, respectively. Two bands cross the Fermi level in each valley. 
		(b) Real-space depiction of valley-protected current across the AB/BA domain boundaries, showing the possible scattering directions for a $K$-valley electron arriving at an AA node in the network. The $K'$ directions are forbidden.
		(c) Depiction of graphene in reciprocal space, including the reciprocal lattice and the Dirac cones at the $K$ and $K'$ points. Inter-valley scattering is indicated by the black arrow.
	}
	\label{fig:TPS}
\end{figure*}

\begin{figure}[t]
	\centering
	\includegraphics[width=0.45\textwidth]{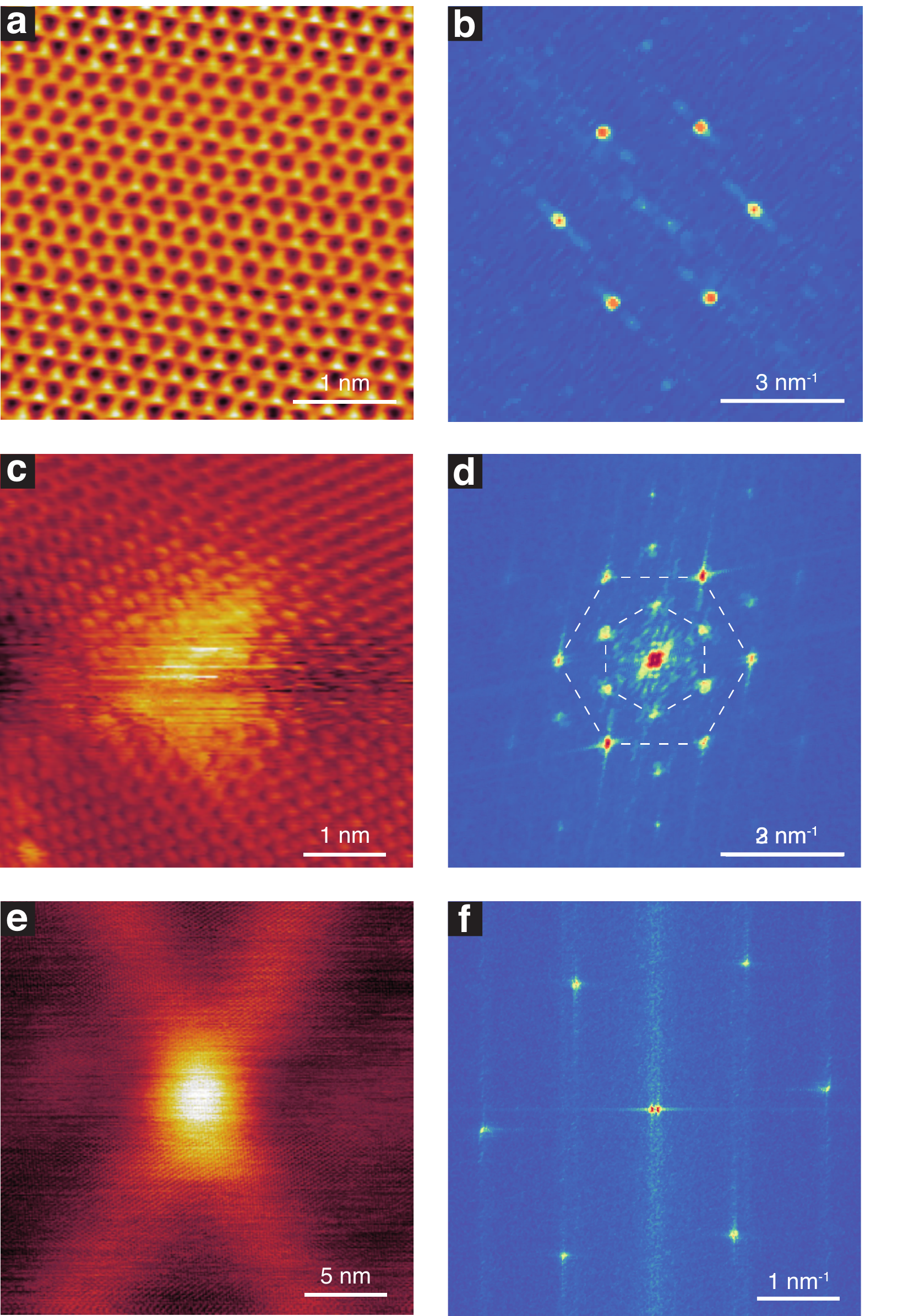}
	\caption{Atomic resolution STM images of HOPG and twisted bilayer graphene including their fast Fourier transforms. 
		(a) STM image of pristine HOPG, showing the honeycomb lattice. The bias voltage is -600 mV and the current setpoint is 1 nA. 
		(b) FFT of (a), showing the 1$\times$1 spots. 
		(c) HOPG surface with an adsorbed hydrogen atom in the center of the image. The bias voltage is -500 mV and the current setpoint is 600 pA. 
		(d) FFT of (c), indicating the presence of the 1$\times$1 spots as well as the ($\surd$3$\times\surd$3)R30\textdegree{} inter-valley scattering spots. White dashed lines indicate the inner and outer hexagonal spots.
		(e) Atomic resolution zoom-in of the STM image of figure~\ref{fig:network}a near one of the AA nodes. The bias voltage is 400 mV and the current setpoint is 500 pA. 
		(f) FFT of (e), showing the graphene 1$\times$1 spots without the presence of inter-valley scattering. }
	\label{fig:STM-FFT}
\end{figure}

The triangular topological network of one-dimensional states has been predicted by theory \cite{Zhang2013,San-Jose2013,Tsim2020,Efimkin2018}, and its existence has been attributed to topological phase transitions across the AB/BA domains inside the moiré pattern. This phase transition is caused by the application of a perpendicular electric field, which breaks the inversion symmetry of the AB and BA stacking configurations. As a result, the valley Chern number of the AB and BA domains is unity with a sign that depends on the stacking configuration and the direction of the electric field. Therefore, at an AB/BA domain boundary, a change in the valley Chern number of $\pm$2 is present. This means that at every boundary there are two topologically protected helical channels per valley, as shown in figure~\ref{fig:TPS}a. Accounting for spin degeneracy, the total conductance of the one-dimensional boundary states is $4e^2/h$. Additionally, the current along the domain boundaries is valley polarized, i.e. the $K$ and $K'$ states counterpropagate \cite{Rickhaus2018,Martin2008,Vaezi2013,Zhang2013,Efimkin2018,Xu2019,Tsim2020}. The topological protection ensures that inter-valley back-scattering is prevented. The valley-protected 1D states span the moiré pattern of the TBG, connecting at the AA stacking regions. This can be interpreted as the domain walls acting as links in the network, whereas the AA domains act as nodes. In the valley-protected network, $K$-states arriving at a node have three valley-preserving scattering directions~\cite{Rickhaus2018,Hou2020}, as shown in figure~\ref{fig:TPS}b. \\
\indent Despite the fact that these topologically protected 1D states have been observed in several studies \cite{Huang2018,Rickhaus2018,Xu2019}, compelling experimental evidence that these 1D states are indeed valley protected is still lacking. The experimental proof for the existence of this valley protection is far from trivial as it requires a completely defect- and impurity-free sample. In this work we will analyze fast Fourier transforms (FFT) of scanning tunneling microscopy (STM) images of small-angle twisted bilayer graphene completely free of defects and impurities. Our experiments reveal that inter-valley scattering in the triangular network of 1D states is fully absent. The latter implies that at an AA node each channel can only propagate in three out of the six possible directions and that these channels can persist over longer length scales, as long the top and bottom graphene layer are free of defects and impurities. \\
\indent The measurements were performed on highly oriented pyrolithic graphite (HOPG) samples supplied by HQ Graphene. HOPG is naturally Bernal stacked, but also frequently contains domains where the top layer is twisted with respect to the underlying graphite, creating a twisted bilayer. All STM data was recorded at room temperature using an Omicron scanning tunneling microscope (STM1) using electrochemically etched tungsten tips. The STM is mounted in an ultra-high vacuum chamber with a base pressure of 3$\times10^{-11}$ mbar. \\
By using scanning tunnelling microscopy and applying a fast Fourier transform to the extracted topography, one can find details inside both the topography and the electronic properties sometimes invisible to even the trained eye. This FFT-STM technique can be applied to graphene, graphite and also bilayer graphene to easily verify the presence or absence of inter-valley scattering. It has previously been demonstrated that inter-valley scattering can be induced in monolayer and bilayer graphene by a defect, such as an atomic vacancy or an adsorbed adatom \cite{Rutter2007,Dutreix2019,Zhang2020}. Figure~\ref{fig:TPS}c shows a schematic of the 1$\times$1 unit cell in reciprocal space (green), as well as the valleys (blue for $K$, red for $K'$) of the band structure of graphene. In reciprocal space, the graphene lattice results in a hexagonal pattern of spots, as indicated in green. In the case of defect-free, free-standing graphene, these spots will appear around the central spot with a vector length of  $\mid$\textbf{b}$\mid = 4\pi/3a \approx$ 2.95 \AA\textsuperscript{-1}. However, when a defect is present in graphene, inter-valley scattering of incoming electron waves will occur at the defect. The latter will result in another set of spots in reciprocal space with a ($\surd$3$\times\surd$3)R30\textdegree{} structure, corresponding to the $K$ and $K'$ valleys, with an associated vector length $\mid$\textbf{$\Delta$K}$\mid = 4\pi/3\sqrt{3}a \approx$ 1.70 \AA\textsuperscript{-1}.\\
To demonstrate this, the Fourier transform of pristine HOPG can be compared to the Fourier transform of HOPG with a surface defect. In figure \ref{fig:STM-FFT}a and b both the real-space STM topography and the FFT of the topography are shown for pristine HOPG. The FFT shows six spots belonging the the 1$\times$1 structure of the honeycomb lattice of graphene. However, when a hydrogen atom is adsorbed on the HOPG, as shown in figure~\ref{fig:STM-FFT}c, the FFT of figure~\ref{fig:STM-FFT}d not only includes the 1$\times$1 spots, but also a ($\surd$3$\times\surd$3)R30\textdegree{} pattern. These spots are due to inter-valley scattering of an incoming electron wave from a $K$ to a $K'$ valley or vice versa.\\
\indent To confirm the absence of inter-valley scattering in twisted bilayer graphene, STM topography images not only have to show the topological network, but also have to be of atomic resolution.  As mentioned previously, the network of topologically protected states emerges when a perpendicular electric field breaks the inversion symmetry between the AB/BA domains. In STM measurements, this electric field is caused by the sample-tip bias $V$ and the difference in work function $\phi$ between the tip and the sample. Furthermore, the application of a back gate also provides an additional factor contributing to the electric field, allowing the gap of the AB/BA regions to be tuned \cite{Huang2015}. The work function difference between sample and tip is difficult to control experimentally, since it is dependent on the tip geometry: however, a minor difference in work function of a few tens of eV already leads to a substantial electric field, given the small sample-tip distances. The latter implies that the topologically protected states can be observed even at zero sample bias: scanning tunnelling spectroscopy (STS) reveals that the local density of states (LDOS) at the domain walls is higher than the AB/BA domain walls at zero bias voltage~\cite{Yao2020}. Figure~\ref{fig:STM-FFT}e shows an atomic resolution image of one of the AA stacked domains from figure \ref{fig:network}a. In this image, also parts of the domain wall states are visible. When taking the FFT of this image it results in the familiar graphene 1$\times$1 structure as seen in figure~\ref{fig:STM-FFT}f. In the FFT, no spots corresponding to ($\surd$3$\times\surd$3)R30\textdegree{} inter-valley scattering are present. This indicates that in the 1D channels, as well as at the AA nodes, the electronic transport is valley protected. \\
\indent This series of experiments provides solid experimental evidence that inside the network of one-dimensional states, no native inter-valley scattering occurs. We argue however, as predicted by theory~\cite{San-Jose2013}, that even a single imperfection, either in the atomic lattice of the top or the bottom layer, will introduce inter-valley scattering. This will locally destroy the topological network and leave the TBG in the gapped phase. In the case of a hydrogen adatom, this can normally be solved by annealing the graphene. However, since previous reports have mentioned that at elevated temperatures TBG is able to 'untwist'~\cite{Wang2015,Kim2017,Cao2018,Kerelsky2019}, reverting to the energetically favoured Bernal stacking, thermally annealing the sample to desorb adatoms should be avoided. For the same reasoning, lattice defects might be more difficult to remove. It has been theoretically predicted that atomic hydrogen preferably chemisorbs at AA-stacked regions in twisted bilayer graphene \cite{Brihuega2018}. This could prove troublesome since this could possibly eliminate an entire node from the network. In contrast, were hydrogen to adsorb at a domain wall, only a single link would be eliminated, leaving more of the network intact.
It is therefore crucial  that if devices are fabricated with the intention of making use of the network of one-dimensional states, the graphene used for such devices should be of very high quality. In addition, a capping layer of for instance hexagonal boron nitride (hBN) could prevent the adsorption of unwanted atoms or molecules to the graphene surface. \\
\indent In conclusion, we have provided compelling experimental evidence that the electronic transport in small-angle twisted bilayer graphene is valley protected, as predicted by theory. By analyzing FFT-STM data of defect-free twisted bilayer graphene, it can be seen that the topologically protected stats are present even at room temperature and that inter-valley scattering is fully absent within the triangular network. This solidifies the prediction that the one-dimensional channels can persist over multiple moiré wavelengths as long as both layers of the twisted bilayer are free of defects.

\newpage
\begin{acknowledgments}
This work was part of the research program on 2D semiconductor crystals with Project No. FV157-TWOD, which was financed by the Netherlands Organization for Scientific Research (NWO). Q.Y. thanks the China Scholarship Council for financial support.
\end{acknowledgments}


\begin{thebibliography}{32}%
	\makeatletter
	\providecommand \@ifxundefined [1]{%
		\@ifx{#1\undefined}
	}%
	\providecommand \@ifnum [1]{%
		\ifnum #1\expandafter \@firstoftwo
		\else \expandafter \@secondoftwo
		\fi
	}%
	\providecommand \@ifx [1]{%
		\ifx #1\expandafter \@firstoftwo
		\else \expandafter \@secondoftwo
		\fi
	}%
	\providecommand \natexlab [1]{#1}%
	\providecommand \enquote  [1]{``#1''}%
	\providecommand \bibnamefont  [1]{#1}%
	\providecommand \bibfnamefont [1]{#1}%
	\providecommand \citenamefont [1]{#1}%
	\providecommand \href@noop [0]{\@secondoftwo}%
	\providecommand \href [0]{\begingroup \@sanitize@url \@href}%
	\providecommand \@href[1]{\@@startlink{#1}\@@href}%
	\providecommand \@@href[1]{\endgroup#1\@@endlink}%
	\providecommand \@sanitize@url [0]{\catcode `\\12\catcode `\$12\catcode
		`\&12\catcode `\#12\catcode `\^12\catcode `\_12\catcode `\%12\relax}%
	\providecommand \@@startlink[1]{}%
	\providecommand \@@endlink[0]{}%
	\providecommand \url  [0]{\begingroup\@sanitize@url \@url }%
	\providecommand \@url [1]{\endgroup\@href {#1}{\urlprefix }}%
	\providecommand \urlprefix  [0]{URL }%
	\providecommand \Eprint [0]{\href }%
	\providecommand \doibase [0]{https://doi.org/}%
	\providecommand \selectlanguage [0]{\@gobble}%
	\providecommand \bibinfo  [0]{\@secondoftwo}%
	\providecommand \bibfield  [0]{\@secondoftwo}%
	\providecommand \translation [1]{[#1]}%
	\providecommand \BibitemOpen [0]{}%
	\providecommand \bibitemStop [0]{}%
	\providecommand \bibitemNoStop [0]{.\EOS\space}%
	\providecommand \EOS [0]{\spacefactor3000\relax}%
	\providecommand \BibitemShut  [1]{\csname bibitem#1\endcsname}%
	\let\auto@bib@innerbib\@empty
	\bibitem [{\citenamefont {Novoselov}\ \emph {et~al.}(2004)\citenamefont
		{Novoselov}, \citenamefont {Geim}, \citenamefont {Morozov}, \citenamefont
		{Jiang}, \citenamefont {Zhang}, \citenamefont {Dubonos}, \citenamefont
		{Grigorieva},\ and\ \citenamefont {Firsov}}]{Novoselov2004}%
	\BibitemOpen
	\bibfield  {author} {\bibinfo {author} {\bibfnamefont {K.~S.}\ \bibnamefont
			{Novoselov}}, \bibinfo {author} {\bibfnamefont {A.~K.}\ \bibnamefont {Geim}},
		\bibinfo {author} {\bibfnamefont {S.~V.}\ \bibnamefont {Morozov}}, \bibinfo
		{author} {\bibfnamefont {D.}~\bibnamefont {Jiang}}, \bibinfo {author}
		{\bibfnamefont {Y.}~\bibnamefont {Zhang}}, \bibinfo {author} {\bibfnamefont
			{S.~V.}\ \bibnamefont {Dubonos}}, \bibinfo {author} {\bibfnamefont {I.~V.}\
			\bibnamefont {Grigorieva}},\ and\ \bibinfo {author} {\bibfnamefont {A.~A.}\
			\bibnamefont {Firsov}},\ }\bibfield  {title} {\bibinfo {title} {{Electric
				field in atomically thin carbon films}},\ }\href
	{https://doi.org/10.1126/science.1102896} {\bibfield  {journal} {\bibinfo
			{journal} {Science}\ }\textbf {\bibinfo {volume} {306}},\ \bibinfo {pages}
		{666} (\bibinfo {year} {2004})},\ \Eprint {https://arxiv.org/abs/0410550}
	{arXiv:0410550 [cond-mat]} \BibitemShut {NoStop}%
	\bibitem [{\citenamefont {Geim}\ and\ \citenamefont
		{Novoselov}(2007)}]{Geim2007}%
	\BibitemOpen
	\bibfield  {author} {\bibinfo {author} {\bibfnamefont {A.~K.}\ \bibnamefont
			{Geim}}\ and\ \bibinfo {author} {\bibfnamefont {K.~S.}\ \bibnamefont
			{Novoselov}},\ }\bibfield  {title} {\bibinfo {title} {{The rise of
				graphene}},\ }\href {https://doi.org/10.1038/nmat1849} {\bibfield  {journal}
		{\bibinfo  {journal} {Nature Materials}\ }\textbf {\bibinfo {volume} {6}},\
		\bibinfo {pages} {183} (\bibinfo {year} {2007})}\BibitemShut {NoStop}%
	\bibitem [{\citenamefont {Wallace}(1947)}]{Wallace1947}%
	\BibitemOpen
	\bibfield  {author} {\bibinfo {author} {\bibfnamefont {P.~R.}\ \bibnamefont
			{Wallace}},\ }\bibfield  {title} {\bibinfo {title} {{The band theory of
				graphite}},\ }\href {https://doi.org/10.1103/PhysRev.71.622} {\bibfield
		{journal} {\bibinfo  {journal} {Physical Review}\ }\textbf {\bibinfo {volume}
			{71}},\ \bibinfo {pages} {622} (\bibinfo {year} {1947})}\BibitemShut
	{NoStop}%
	\bibitem [{\citenamefont {{Castro Neto}}(2009)}]{Neto2009}%
	\BibitemOpen
	\bibfield  {author} {\bibinfo {author} {\bibfnamefont {A.~H.}\ \bibnamefont
			{{Castro Neto}}},\ }\bibfield  {title} {\bibinfo {title} {{The electronic
				properties of graphene}},\ }\href {https://doi.org/10.1103/RevModPhys.81.109}
	{\bibfield  {journal} {\bibinfo  {journal} {Reviews of Modern Physics}\
		}\textbf {\bibinfo {volume} {81}},\ \bibinfo {pages} {109} (\bibinfo {year}
		{2009})}\BibitemShut {NoStop}%
	\bibitem [{\citenamefont {Li}\ \emph {et~al.}(2010)\citenamefont {Li},
		\citenamefont {Luican}, \citenamefont {{Lopes Dos Santos}}, \citenamefont
		{{Castro Neto}}, \citenamefont {Reina}, \citenamefont {Kong},\ and\
		\citenamefont {Andrei}}]{Li2010}%
	\BibitemOpen
	\bibfield  {author} {\bibinfo {author} {\bibfnamefont {G.}~\bibnamefont
			{Li}}, \bibinfo {author} {\bibfnamefont {A.}~\bibnamefont {Luican}}, \bibinfo
		{author} {\bibfnamefont {J.~M.}\ \bibnamefont {{Lopes Dos Santos}}}, \bibinfo
		{author} {\bibfnamefont {A.~H.}\ \bibnamefont {{Castro Neto}}}, \bibinfo
		{author} {\bibfnamefont {A.}~\bibnamefont {Reina}}, \bibinfo {author}
		{\bibfnamefont {J.}~\bibnamefont {Kong}},\ and\ \bibinfo {author}
		{\bibfnamefont {E.~Y.}\ \bibnamefont {Andrei}},\ }\bibfield  {title}
	{\bibinfo {title} {{Observation of Van Hove singularities in twisted graphene
				layers}},\ }\href {https://doi.org/10.1038/nphys1463} {\bibfield  {journal}
		{\bibinfo  {journal} {Nature Physics}\ }\textbf {\bibinfo {volume} {6}},\
		\bibinfo {pages} {109} (\bibinfo {year} {2010})},\ \Eprint
	{https://arxiv.org/abs/0912.2102} {arXiv:0912.2102} \BibitemShut {NoStop}%
	\bibitem [{\citenamefont {Yan}\ \emph {et~al.}(2012)\citenamefont {Yan},
		\citenamefont {Liu}, \citenamefont {Dou}, \citenamefont {Meng}, \citenamefont
		{Feng}, \citenamefont {Chu}, \citenamefont {Zhang}, \citenamefont {Liu},
		\citenamefont {Nie},\ and\ \citenamefont {He}}]{Yan2012}%
	\BibitemOpen
	\bibfield  {author} {\bibinfo {author} {\bibfnamefont {W.}~\bibnamefont
			{Yan}}, \bibinfo {author} {\bibfnamefont {M.}~\bibnamefont {Liu}}, \bibinfo
		{author} {\bibfnamefont {R.~F.}\ \bibnamefont {Dou}}, \bibinfo {author}
		{\bibfnamefont {L.}~\bibnamefont {Meng}}, \bibinfo {author} {\bibfnamefont
			{L.}~\bibnamefont {Feng}}, \bibinfo {author} {\bibfnamefont {Z.~D.}\
			\bibnamefont {Chu}}, \bibinfo {author} {\bibfnamefont {Y.}~\bibnamefont
			{Zhang}}, \bibinfo {author} {\bibfnamefont {Z.}~\bibnamefont {Liu}}, \bibinfo
		{author} {\bibfnamefont {J.~C.}\ \bibnamefont {Nie}},\ and\ \bibinfo {author}
		{\bibfnamefont {L.}~\bibnamefont {He}},\ }\bibfield  {title} {\bibinfo
		{title} {{Angle-dependent van Hove singularities in a slightly twisted
				graphene bilayer}},\ }\href {https://doi.org/10.1103/PhysRevLett.109.126801}
	{\bibfield  {journal} {\bibinfo  {journal} {Physical Review Letters}\
		}\textbf {\bibinfo {volume} {109}},\ \bibinfo {pages} {1} (\bibinfo {year}
		{2012})},\ \Eprint {https://arxiv.org/abs/1206.5883} {arXiv:1206.5883}
	\BibitemShut {NoStop}%
	\bibitem [{\citenamefont {Brihuega}\ \emph {et~al.}(2012)\citenamefont
		{Brihuega}, \citenamefont {Mallet}, \citenamefont {Gonz{\'{a}}lez-Herrero},
		\citenamefont {{Trambly De Laissardi{\`{e}}}}, \citenamefont {Ugeda},
		\citenamefont {Magaud}, \citenamefont {G{\'{o}}mez-Rodr{\'{i}}guez},
		\citenamefont {Yndur{\'{a}}in},\ and\ \citenamefont
		{Veuillen}}]{Brihuega2012}%
	\BibitemOpen
	\bibfield  {author} {\bibinfo {author} {\bibfnamefont {I.}~\bibnamefont
			{Brihuega}}, \bibinfo {author} {\bibfnamefont {P.}~\bibnamefont {Mallet}},
		\bibinfo {author} {\bibfnamefont {H.}~\bibnamefont {Gonz{\'{a}}lez-Herrero}},
		\bibinfo {author} {\bibfnamefont {G.}~\bibnamefont {{Trambly De
					Laissardi{\`{e}}}}}, \bibinfo {author} {\bibfnamefont {M.~M.}\ \bibnamefont
			{Ugeda}}, \bibinfo {author} {\bibfnamefont {L.}~\bibnamefont {Magaud}},
		\bibinfo {author} {\bibfnamefont {J.~M.}\ \bibnamefont
			{G{\'{o}}mez-Rodr{\'{i}}guez}}, \bibinfo {author} {\bibfnamefont
			{F.}~\bibnamefont {Yndur{\'{a}}in}},\ and\ \bibinfo {author} {\bibfnamefont
			{J.-Y.}\ \bibnamefont {Veuillen}},\ }\bibfield  {title} {\bibinfo {title}
		{{Unraveling the Intrinsic and Robust Nature of van Hove Singularities in
				Twisted Bilayer Graphene by Scanning Tunneling Microscopy and Theoretical
				Analysis}}\ }\href {https://doi.org/10.1103/PhysRevLett.109.196802}
	{10.1103/PhysRevLett.109.196802} (\bibinfo {year} {2012})\BibitemShut
	{NoStop}%
	\bibitem [{\citenamefont {{Van Hove}}(1953)}]{VanHove1953}%
	\BibitemOpen
	\bibfield  {author} {\bibinfo {author} {\bibfnamefont {L.}~\bibnamefont {{Van
					Hove}}},\ }\bibfield  {title} {\bibinfo {title} {{The occurrence of
				singularities in the elastic frequency distribution of a crystal}},\ }\href
	{https://doi.org/10.1103/PhysRev.89.1189} {\bibfield  {journal} {\bibinfo
			{journal} {Physical Review}\ }\textbf {\bibinfo {volume} {89}},\ \bibinfo
		{pages} {1189} (\bibinfo {year} {1953})}\BibitemShut {NoStop}%
	\bibitem [{\citenamefont {Cao}\ \emph {et~al.}(2018)\citenamefont {Cao},
		\citenamefont {Fatemi}, \citenamefont {Fang}, \citenamefont {Watanabe},
		\citenamefont {Taniguchi}, \citenamefont {Kaxiras},\ and\ \citenamefont
		{Jarillo-Herrero}}]{Cao2018}%
	\BibitemOpen
	\bibfield  {author} {\bibinfo {author} {\bibfnamefont {Y.}~\bibnamefont
			{Cao}}, \bibinfo {author} {\bibfnamefont {V.}~\bibnamefont {Fatemi}},
		\bibinfo {author} {\bibfnamefont {S.}~\bibnamefont {Fang}}, \bibinfo {author}
		{\bibfnamefont {K.}~\bibnamefont {Watanabe}}, \bibinfo {author}
		{\bibfnamefont {T.}~\bibnamefont {Taniguchi}}, \bibinfo {author}
		{\bibfnamefont {E.}~\bibnamefont {Kaxiras}},\ and\ \bibinfo {author}
		{\bibfnamefont {P.}~\bibnamefont {Jarillo-Herrero}},\ }\bibfield  {title}
	{\bibinfo {title} {{Unconventional superconductivity in magic-angle graphene
				superlattices}},\ }\href {https://doi.org/10.1038/nature26160} {\bibfield
		{journal} {\bibinfo  {journal} {Nature}\ }\textbf {\bibinfo {volume} {556}},\
		\bibinfo {pages} {43} (\bibinfo {year} {2018})},\ \Eprint
	{https://arxiv.org/abs/1803.02342} {arXiv:1803.02342} \BibitemShut {NoStop}%
	\bibitem [{\citenamefont {Yankowitz}\ \emph {et~al.}(2019)\citenamefont
		{Yankowitz}, \citenamefont {Chen}, \citenamefont {Polshyn}, \citenamefont
		{Zhang}, \citenamefont {Watanabe}, \citenamefont {Taniguchi}, \citenamefont
		{Graf}, \citenamefont {Young},\ and\ \citenamefont {Dean}}]{Yankowitz2019}%
	\BibitemOpen
	\bibfield  {author} {\bibinfo {author} {\bibfnamefont {M.}~\bibnamefont
			{Yankowitz}}, \bibinfo {author} {\bibfnamefont {S.}~\bibnamefont {Chen}},
		\bibinfo {author} {\bibfnamefont {H.}~\bibnamefont {Polshyn}}, \bibinfo
		{author} {\bibfnamefont {Y.}~\bibnamefont {Zhang}}, \bibinfo {author}
		{\bibfnamefont {K.}~\bibnamefont {Watanabe}}, \bibinfo {author}
		{\bibfnamefont {T.}~\bibnamefont {Taniguchi}}, \bibinfo {author}
		{\bibfnamefont {D.}~\bibnamefont {Graf}}, \bibinfo {author} {\bibfnamefont
			{A.~F.}\ \bibnamefont {Young}},\ and\ \bibinfo {author} {\bibfnamefont
			{C.~R.}\ \bibnamefont {Dean}},\ }\bibfield  {title} {\bibinfo {title}
		{{Tuning superconductivity in twisted bilayer graphene}},\ }\href
	{https://doi.org/10.1126/science.aav1910} {\bibfield  {journal} {\bibinfo
			{journal} {Science}\ }\textbf {\bibinfo {volume} {363}},\ \bibinfo {pages}
		{1059} (\bibinfo {year} {2019})}\BibitemShut {NoStop}%
	\bibitem [{\citenamefont {Padhi}\ \emph {et~al.}(2018)\citenamefont {Padhi},
		\citenamefont {Setty},\ and\ \citenamefont {Phillips}}]{Padhi2018}%
	\BibitemOpen
	\bibfield  {author} {\bibinfo {author} {\bibfnamefont {B.}~\bibnamefont
			{Padhi}}, \bibinfo {author} {\bibfnamefont {C.}~\bibnamefont {Setty}},\ and\
		\bibinfo {author} {\bibfnamefont {P.~W.}\ \bibnamefont {Phillips}},\
	}\bibfield  {title} {\bibinfo {title} {{Doped Twisted Bilayer Graphene near
				Magic Angles: Proximity to Wigner Crystallization, Not Mott Insulation}}\
	}\href {https://doi.org/10.1021/acs.nanolett.8b02033}
	{10.1021/acs.nanolett.8b02033} (\bibinfo {year} {2018})\BibitemShut {NoStop}%
	\bibitem [{\citenamefont {Sharpe}\ \emph {et~al.}(2019)\citenamefont {Sharpe},
		\citenamefont {Fox}, \citenamefont {Barnard}, \citenamefont {Finney},
		\citenamefont {Watanabe}, \citenamefont {Taniguchi}, \citenamefont
		{Kastner},\ and\ \citenamefont {Goldhaber-Gordon}}]{Sharpe2019}%
	\BibitemOpen
	\bibfield  {author} {\bibinfo {author} {\bibfnamefont {A.~L.}\ \bibnamefont
			{Sharpe}}, \bibinfo {author} {\bibfnamefont {E.~J.}\ \bibnamefont {Fox}},
		\bibinfo {author} {\bibfnamefont {A.~W.}\ \bibnamefont {Barnard}}, \bibinfo
		{author} {\bibfnamefont {J.}~\bibnamefont {Finney}}, \bibinfo {author}
		{\bibfnamefont {K.}~\bibnamefont {Watanabe}}, \bibinfo {author}
		{\bibfnamefont {T.}~\bibnamefont {Taniguchi}}, \bibinfo {author}
		{\bibfnamefont {M.~A.}\ \bibnamefont {Kastner}},\ and\ \bibinfo {author}
		{\bibfnamefont {D.}~\bibnamefont {Goldhaber-Gordon}},\ }\bibfield  {title}
	{\bibinfo {title} {{Emergent ferromagnetism near three-quarters filling in
				twisted bilayer graphene.}},\ }\href
	{https://doi.org/10.1126/science.aaw3780} {\bibfield  {journal} {\bibinfo
			{journal} {Science (New York, N.Y.)}\ }\textbf {\bibinfo {volume} {365}},\
		\bibinfo {pages} {605} (\bibinfo {year} {2019})}\BibitemShut {NoStop}%
	\bibitem [{\citenamefont {Gonzalez-Arraga}\ \emph {et~al.}(2017)\citenamefont
		{Gonzalez-Arraga}, \citenamefont {Lado}, \citenamefont {Guinea},\ and\
		\citenamefont {San-Jose}}]{Gonzalez-Arraga2017}%
	\BibitemOpen
	\bibfield  {author} {\bibinfo {author} {\bibfnamefont {L.~A.}\ \bibnamefont
			{Gonzalez-Arraga}}, \bibinfo {author} {\bibfnamefont {J.~L.}\ \bibnamefont
			{Lado}}, \bibinfo {author} {\bibfnamefont {F.}~\bibnamefont {Guinea}},\ and\
		\bibinfo {author} {\bibfnamefont {P.}~\bibnamefont {San-Jose}},\ }\bibfield
	{title} {\bibinfo {title} {{Electrically Controllable Magnetism in Twisted
				Bilayer Graphene}},\ }\href {https://doi.org/10.1103/PhysRevLett.119.107201}
	{\bibfield  {journal} {\bibinfo  {journal} {Physical Review Letters}\
		}\textbf {\bibinfo {volume} {119}},\ \bibinfo {pages} {107201} (\bibinfo
		{year} {2017})},\ \Eprint {https://arxiv.org/abs/1702.08831}
	{arXiv:1702.08831} \BibitemShut {NoStop}%
	\bibitem [{\citenamefont {Huang}\ \emph {et~al.}(2018)\citenamefont {Huang},
		\citenamefont {Kim}, \citenamefont {Efimkin}, \citenamefont {Lovorn},
		\citenamefont {Taniguchi}, \citenamefont {Watanabe}, \citenamefont
		{Macdonald}, \citenamefont {Tutuc},\ and\ \citenamefont {Leroy}}]{Huang2018}%
	\BibitemOpen
	\bibfield  {author} {\bibinfo {author} {\bibfnamefont {S.}~\bibnamefont
			{Huang}}, \bibinfo {author} {\bibfnamefont {K.}~\bibnamefont {Kim}}, \bibinfo
		{author} {\bibfnamefont {D.~K.}\ \bibnamefont {Efimkin}}, \bibinfo {author}
		{\bibfnamefont {T.}~\bibnamefont {Lovorn}}, \bibinfo {author} {\bibfnamefont
			{T.}~\bibnamefont {Taniguchi}}, \bibinfo {author} {\bibfnamefont
			{K.}~\bibnamefont {Watanabe}}, \bibinfo {author} {\bibfnamefont {A.~H.}\
			\bibnamefont {Macdonald}}, \bibinfo {author} {\bibfnamefont {E.}~\bibnamefont
			{Tutuc}},\ and\ \bibinfo {author} {\bibfnamefont {B.~J.}\ \bibnamefont
			{Leroy}},\ }\bibfield  {title} {\bibinfo {title} {{Topologically Protected
				Helical States in Minimally Twisted Bilayer Graphene}},\ }\href
	{https://doi.org/10.1103/PhysRevLett.121.037702} {\bibfield  {journal}
		{\bibinfo  {journal} {Physical Review Letters}\ }\textbf {\bibinfo {volume}
			{121}},\ \bibinfo {pages} {37702} (\bibinfo {year} {2018})}\BibitemShut
	{NoStop}%
	\bibitem [{\citenamefont {Rickhaus}\ \emph {et~al.}(2018)\citenamefont
		{Rickhaus}, \citenamefont {Wallbank}, \citenamefont {Slizovskiy},
		\citenamefont {Pisoni}, \citenamefont {Overweg}, \citenamefont {Lee},
		\citenamefont {Eich}, \citenamefont {Liu}, \citenamefont {Watanabe},
		\citenamefont {Taniguchi}, \citenamefont {Ihn},\ and\ \citenamefont
		{Ensslin}}]{Rickhaus2018}%
	\BibitemOpen
	\bibfield  {author} {\bibinfo {author} {\bibfnamefont {P.}~\bibnamefont
			{Rickhaus}}, \bibinfo {author} {\bibfnamefont {J.}~\bibnamefont {Wallbank}},
		\bibinfo {author} {\bibfnamefont {S.}~\bibnamefont {Slizovskiy}}, \bibinfo
		{author} {\bibfnamefont {R.}~\bibnamefont {Pisoni}}, \bibinfo {author}
		{\bibfnamefont {H.}~\bibnamefont {Overweg}}, \bibinfo {author} {\bibfnamefont
			{Y.}~\bibnamefont {Lee}}, \bibinfo {author} {\bibfnamefont {M.}~\bibnamefont
			{Eich}}, \bibinfo {author} {\bibfnamefont {M.-H.}\ \bibnamefont {Liu}},
		\bibinfo {author} {\bibfnamefont {K.}~\bibnamefont {Watanabe}}, \bibinfo
		{author} {\bibfnamefont {T.}~\bibnamefont {Taniguchi}}, \bibinfo {author}
		{\bibfnamefont {T.}~\bibnamefont {Ihn}},\ and\ \bibinfo {author}
		{\bibfnamefont {K.}~\bibnamefont {Ensslin}},\ }\bibfield  {title} {\bibinfo
		{title} {{Transport Through a Network of Topological Channels in Twisted
				Bilayer Graphene}},\ }\href {https://doi.org/10.1021/acs.nanolett.8b02387}
	{\bibfield  {journal} {\bibinfo  {journal} {Nano Lett}\ }\textbf {\bibinfo
			{volume} {18}},\ \bibinfo {pages} {47} (\bibinfo {year} {2018})}\BibitemShut
	{NoStop}%
	\bibitem [{\citenamefont {Xu}\ \emph {et~al.}(2019)\citenamefont {Xu},
		\citenamefont {Berdyugin}, \citenamefont {Kumaravadivel}, \citenamefont
		{Guinea}, \citenamefont {{Krishna Kumar}}, \citenamefont {Bandurin},
		\citenamefont {Morozov}, \citenamefont {Kuang}, \citenamefont {Tsim},
		\citenamefont {Liu}, \citenamefont {Edgar}, \citenamefont {Grigorieva},
		\citenamefont {Fal'ko}, \citenamefont {Kim},\ and\ \citenamefont
		{Geim}}]{Xu2019}%
	\BibitemOpen
	\bibfield  {author} {\bibinfo {author} {\bibfnamefont {S.~G.}\ \bibnamefont
			{Xu}}, \bibinfo {author} {\bibfnamefont {A.~I.}\ \bibnamefont {Berdyugin}},
		\bibinfo {author} {\bibfnamefont {P.}~\bibnamefont {Kumaravadivel}}, \bibinfo
		{author} {\bibfnamefont {F.}~\bibnamefont {Guinea}}, \bibinfo {author}
		{\bibfnamefont {R.}~\bibnamefont {{Krishna Kumar}}}, \bibinfo {author}
		{\bibfnamefont {D.~A.}\ \bibnamefont {Bandurin}}, \bibinfo {author}
		{\bibfnamefont {S.~V.}\ \bibnamefont {Morozov}}, \bibinfo {author}
		{\bibfnamefont {W.}~\bibnamefont {Kuang}}, \bibinfo {author} {\bibfnamefont
			{B.}~\bibnamefont {Tsim}}, \bibinfo {author} {\bibfnamefont {S.}~\bibnamefont
			{Liu}}, \bibinfo {author} {\bibfnamefont {J.~H.}\ \bibnamefont {Edgar}},
		\bibinfo {author} {\bibfnamefont {I.~V.}\ \bibnamefont {Grigorieva}},
		\bibinfo {author} {\bibfnamefont {V.~I.}\ \bibnamefont {Fal'ko}}, \bibinfo
		{author} {\bibfnamefont {M.}~\bibnamefont {Kim}},\ and\ \bibinfo {author}
		{\bibfnamefont {A.~K.}\ \bibnamefont {Geim}},\ }\bibfield  {title} {\bibinfo
		{title} {{Giant oscillations in a triangular network of one-dimensional
				states in marginally twisted graphene}},\ }\href
	{https://doi.org/10.1038/s41467-019-11971-7} {\bibfield  {journal} {\bibinfo
			{journal} {Nature Communications}\ }\textbf {\bibinfo {volume} {10}},\
		\bibinfo {pages} {3} (\bibinfo {year} {2019})},\ \Eprint
	{https://arxiv.org/abs/1905.12984} {arXiv:1905.12984} \BibitemShut {NoStop}%
	\bibitem [{\citenamefont {Zhang}\ \emph {et~al.}(2013)\citenamefont {Zhang},
		\citenamefont {MacDonald},\ and\ \citenamefont {Mele}}]{Zhang2013}%
	\BibitemOpen
	\bibfield  {author} {\bibinfo {author} {\bibfnamefont {F.}~\bibnamefont
			{Zhang}}, \bibinfo {author} {\bibfnamefont {A.~H.}\ \bibnamefont
			{MacDonald}},\ and\ \bibinfo {author} {\bibfnamefont {E.~J.}\ \bibnamefont
			{Mele}},\ }\bibfield  {title} {\bibinfo {title} {{Valley Chern numbers and
				boundary modes in gapped bilayer graphene}},\ }\href
	{https://doi.org/10.1073/pnas.1308853110} {\bibfield  {journal} {\bibinfo
			{journal} {Proceedings of the National Academy of Sciences of the United
				States of America}\ }\textbf {\bibinfo {volume} {110}},\ \bibinfo {pages}
		{10546} (\bibinfo {year} {2013})}\BibitemShut {NoStop}%
	\bibitem [{\citenamefont {San-Jose}\ and\ \citenamefont
		{Prada}(2013)}]{San-Jose2013}%
	\BibitemOpen
	\bibfield  {author} {\bibinfo {author} {\bibfnamefont {P.}~\bibnamefont
			{San-Jose}}\ and\ \bibinfo {author} {\bibfnamefont {E.}~\bibnamefont
			{Prada}},\ }\bibfield  {title} {\bibinfo {title} {{Helical networks in
				twisted bilayer graphene under interlayer bias}},\ }\href
	{https://doi.org/10.1103/PhysRevB.88.121408} {\bibfield  {journal} {\bibinfo
			{journal} {Physical Review B - Condensed Matter and Materials Physics}\
		}\textbf {\bibinfo {volume} {88}},\ \bibinfo {pages} {1} (\bibinfo {year}
		{2013})}\BibitemShut {NoStop}%
	\bibitem [{\citenamefont {Tsim}\ \emph {et~al.}(2020)\citenamefont {Tsim},
		\citenamefont {Nam},\ and\ \citenamefont {Koshino}}]{Tsim2020}%
	\BibitemOpen
	\bibfield  {author} {\bibinfo {author} {\bibfnamefont {B.}~\bibnamefont
			{Tsim}}, \bibinfo {author} {\bibfnamefont {N.~N.~T.}\ \bibnamefont {Nam}},\
		and\ \bibinfo {author} {\bibfnamefont {M.}~\bibnamefont {Koshino}},\
	}\bibfield  {title} {\bibinfo {title} {{Perfect one-dimensional chiral states
				in biased twisted bilayer graphene}},\ }\href
	{https://doi.org/10.1103/physrevb.101.125409} {\bibfield  {journal} {\bibinfo
			{journal} {Physical Review B}\ }\textbf {\bibinfo {volume} {101}},\ \bibinfo
		{pages} {1} (\bibinfo {year} {2020})},\ \Eprint
	{https://arxiv.org/abs/2001.06257} {arXiv:2001.06257} \BibitemShut {NoStop}%
	\bibitem [{\citenamefont {Efimkin}\ and\ \citenamefont
		{Macdonald}(2018)}]{Efimkin2018}%
	\BibitemOpen
	\bibfield  {author} {\bibinfo {author} {\bibfnamefont {D.~K.}\ \bibnamefont
			{Efimkin}}\ and\ \bibinfo {author} {\bibfnamefont {A.~H.}\ \bibnamefont
			{Macdonald}},\ }\bibfield  {title} {\bibinfo {title} {{Helical network model
				for twisted bilayer graphene}},\ }\href
	{https://doi.org/10.1103/PhysRevB.98.035404} {\bibfield  {journal} {\bibinfo
			{journal} {Physical Review B}\ }\textbf {\bibinfo {volume} {98}},\ \bibinfo
		{pages} {1} (\bibinfo {year} {2018})},\ \Eprint
	{https://arxiv.org/abs/1803.06404} {arXiv:1803.06404} \BibitemShut {NoStop}%
	\bibitem [{\citenamefont {Martin}\ \emph {et~al.}(2008)\citenamefont {Martin},
		\citenamefont {Blanter},\ and\ \citenamefont {Morpurgo}}]{Martin2008}%
	\BibitemOpen
	\bibfield  {author} {\bibinfo {author} {\bibfnamefont {I.}~\bibnamefont
			{Martin}}, \bibinfo {author} {\bibfnamefont {Y.~M.}\ \bibnamefont
			{Blanter}},\ and\ \bibinfo {author} {\bibfnamefont {A.~F.}\ \bibnamefont
			{Morpurgo}},\ }\bibfield  {title} {\bibinfo {title} {{Topological confinement
				in bilayer graphene}},\ }\href
	{https://doi.org/10.1103/PhysRevLett.100.036804} {\bibfield  {journal}
		{\bibinfo  {journal} {Physical Review Letters}\ }\textbf {\bibinfo {volume}
			{100}},\ \bibinfo {pages} {1} (\bibinfo {year} {2008})},\ \Eprint
	{https://arxiv.org/abs/0709.3522} {arXiv:0709.3522} \BibitemShut {NoStop}%
	\bibitem [{\citenamefont {Vaezi}\ \emph {et~al.}(2013)\citenamefont {Vaezi},
		\citenamefont {Liang}, \citenamefont {Ngai}, \citenamefont {Yang},\ and\
		\citenamefont {Kim}}]{Vaezi2013}%
	\BibitemOpen
	\bibfield  {author} {\bibinfo {author} {\bibfnamefont {A.}~\bibnamefont
			{Vaezi}}, \bibinfo {author} {\bibfnamefont {Y.}~\bibnamefont {Liang}},
		\bibinfo {author} {\bibfnamefont {D.~H.}\ \bibnamefont {Ngai}}, \bibinfo
		{author} {\bibfnamefont {L.}~\bibnamefont {Yang}},\ and\ \bibinfo {author}
		{\bibfnamefont {E.~A.}\ \bibnamefont {Kim}},\ }\bibfield  {title} {\bibinfo
		{title} {{Topological edge states at a tilt boundary in gated multilayer
				graphene}},\ }\href {https://doi.org/10.1103/PhysRevX.3.021018} {\bibfield
		{journal} {\bibinfo  {journal} {Physical Review X}\ }\textbf {\bibinfo
			{volume} {3}},\ \bibinfo {pages} {1} (\bibinfo {year} {2013})}\BibitemShut
	{NoStop}%
	\bibitem [{\citenamefont {Hou}\ \emph {et~al.}(2020)\citenamefont {Hou},
		\citenamefont {Ren}, \citenamefont {Quan}, \citenamefont {Jung},
		\citenamefont {Ren},\ and\ \citenamefont {Qiao}}]{Hou2020}%
	\BibitemOpen
	\bibfield  {author} {\bibinfo {author} {\bibfnamefont {T.}~\bibnamefont
			{Hou}}, \bibinfo {author} {\bibfnamefont {Y.}~\bibnamefont {Ren}}, \bibinfo
		{author} {\bibfnamefont {Y.}~\bibnamefont {Quan}}, \bibinfo {author}
		{\bibfnamefont {J.}~\bibnamefont {Jung}}, \bibinfo {author} {\bibfnamefont
			{W.}~\bibnamefont {Ren}},\ and\ \bibinfo {author} {\bibfnamefont
			{Z.}~\bibnamefont {Qiao}},\ }\bibfield  {title} {\bibinfo {title}
		{{Valley-Current Splitter in Minimally Twisted Bilayer Graphene}},\ }\href
	{https://doi.org/10.1103/PhysRevB.102.085433} {\bibfield  {journal} {\bibinfo
			{journal} {Physical Review B}\ }\textbf {\bibinfo {volume} {102}},\ \bibinfo
		{pages} {1} (\bibinfo {year} {2020})}\BibitemShut {NoStop}%
	\bibitem [{\citenamefont {Rutter}\ \emph {et~al.}(2007)\citenamefont {Rutter},
		\citenamefont {Crain}, \citenamefont {Guisinger}, \citenamefont {Li},
		\citenamefont {First},\ and\ \citenamefont {Stroscio}}]{Rutter2007}%
	\BibitemOpen
	\bibfield  {author} {\bibinfo {author} {\bibfnamefont {G.~M.}\ \bibnamefont
			{Rutter}}, \bibinfo {author} {\bibfnamefont {J.~N.}\ \bibnamefont {Crain}},
		\bibinfo {author} {\bibfnamefont {N.~P.}\ \bibnamefont {Guisinger}}, \bibinfo
		{author} {\bibfnamefont {T.}~\bibnamefont {Li}}, \bibinfo {author}
		{\bibfnamefont {P.~N.}\ \bibnamefont {First}},\ and\ \bibinfo {author}
		{\bibfnamefont {J.~A.}\ \bibnamefont {Stroscio}},\ }\bibfield  {title}
	{\bibinfo {title} {{Scattering and interference in epitaxial graphene}},\
	}\href {https://doi.org/10.1126/science.1142882} {\bibfield  {journal}
		{\bibinfo  {journal} {Science}\ }\textbf {\bibinfo {volume} {317}},\ \bibinfo
		{pages} {219} (\bibinfo {year} {2007})}\BibitemShut {NoStop}%
	\bibitem [{\citenamefont {Dutreix}\ \emph {et~al.}(2019)\citenamefont
		{Dutreix}, \citenamefont {Gonz{\'{a}}lez-Herrero}, \citenamefont {Brihuega},
		\citenamefont {Katsnelson}, \citenamefont {Chapelier},\ and\ \citenamefont
		{Renard}}]{Dutreix2019}%
	\BibitemOpen
	\bibfield  {author} {\bibinfo {author} {\bibfnamefont {C.}~\bibnamefont
			{Dutreix}}, \bibinfo {author} {\bibfnamefont {H.}~\bibnamefont
			{Gonz{\'{a}}lez-Herrero}}, \bibinfo {author} {\bibfnamefont {I.}~\bibnamefont
			{Brihuega}}, \bibinfo {author} {\bibfnamefont {M.~I.}\ \bibnamefont
			{Katsnelson}}, \bibinfo {author} {\bibfnamefont {C.}~\bibnamefont
			{Chapelier}},\ and\ \bibinfo {author} {\bibfnamefont {V.~T.}\ \bibnamefont
			{Renard}},\ }\bibfield  {title} {\bibinfo {title} {{Measuring the Berry phase
				of graphene from wavefront dislocations in Friedel oscillations}},\ }\href
	{https://doi.org/10.1038/s41586-019-1613-5} {\bibfield  {journal} {\bibinfo
			{journal} {Nature}\ }\textbf {\bibinfo {volume} {574}},\ \bibinfo {pages}
		{219} (\bibinfo {year} {2019})},\ \Eprint {https://arxiv.org/abs/1910.00437}
	{arXiv:1910.00437} \BibitemShut {NoStop}%
	\bibitem [{\citenamefont {Zhang}\ \emph {et~al.}(2020)\citenamefont {Zhang},
		\citenamefont {Wang}, \citenamefont {Watanabe}, \citenamefont {Taniguchi},
		\citenamefont {Ueno}, \citenamefont {Tutuc},\ and\ \citenamefont
		{LeRoy}}]{Zhang2020}%
	\BibitemOpen
	\bibfield  {author} {\bibinfo {author} {\bibfnamefont {Z.}~\bibnamefont
			{Zhang}}, \bibinfo {author} {\bibfnamefont {Y.}~\bibnamefont {Wang}},
		\bibinfo {author} {\bibfnamefont {K.}~\bibnamefont {Watanabe}}, \bibinfo
		{author} {\bibfnamefont {T.}~\bibnamefont {Taniguchi}}, \bibinfo {author}
		{\bibfnamefont {K.}~\bibnamefont {Ueno}}, \bibinfo {author} {\bibfnamefont
			{E.}~\bibnamefont {Tutuc}},\ and\ \bibinfo {author} {\bibfnamefont {B.~J.}\
			\bibnamefont {LeRoy}},\ }\bibfield  {title} {\bibinfo {title} {{Flat bands in
				twisted bilayer transition metal dichalcogenides}},\ }\bibfield  {journal}
	{\bibinfo  {journal} {Nature Physics}\ }\href
	{https://doi.org/10.1038/s41567-020-0958-x} {10.1038/s41567-020-0958-x}
	(\bibinfo {year} {2020})\BibitemShut {NoStop}%
	\bibitem [{\citenamefont {Huang}\ \emph {et~al.}(2015)\citenamefont {Huang},
		\citenamefont {Sutter}, \citenamefont {Shi}, \citenamefont {Zheng},
		\citenamefont {Yang}, \citenamefont {Englund}, \citenamefont {Gao},\ and\
		\citenamefont {Sutter}}]{Huang2015}%
	\BibitemOpen
	\bibfield  {author} {\bibinfo {author} {\bibfnamefont {Y.}~\bibnamefont
			{Huang}}, \bibinfo {author} {\bibfnamefont {E.}~\bibnamefont {Sutter}},
		\bibinfo {author} {\bibfnamefont {N.~N.}\ \bibnamefont {Shi}}, \bibinfo
		{author} {\bibfnamefont {J.}~\bibnamefont {Zheng}}, \bibinfo {author}
		{\bibfnamefont {T.}~\bibnamefont {Yang}}, \bibinfo {author} {\bibfnamefont
			{D.}~\bibnamefont {Englund}}, \bibinfo {author} {\bibfnamefont {H.-J.}\
			\bibnamefont {Gao}},\ and\ \bibinfo {author} {\bibfnamefont {P.}~\bibnamefont
			{Sutter}},\ }\bibfield  {title} {\bibinfo {title} {{Reliable Exfoliation of
				Large-Area High-Quality Flakes of Graphene and Other Two-Dimensional
				Materials}},\ }\href {https://doi.org/10.1021/acsnano.5b04258} {\bibfield
		{journal} {\bibinfo  {journal} {ACS Nano}\ }\textbf {\bibinfo {volume} {9}},\
		\bibinfo {pages} {10612} (\bibinfo {year} {2015})}\BibitemShut {NoStop}%
	\bibitem [{\citenamefont {Yao}\ \emph {et~al.}(2020)\citenamefont {Yao},
		\citenamefont {Chen}, \citenamefont {{Van Bremen}}, \citenamefont
		{Sotthewes},\ and\ \citenamefont {Zandvliet}}]{Yao2020}%
	\BibitemOpen
	\bibfield  {author} {\bibinfo {author} {\bibfnamefont {Q.}~\bibnamefont
			{Yao}}, \bibinfo {author} {\bibfnamefont {X.}~\bibnamefont {Chen}}, \bibinfo
		{author} {\bibfnamefont {R.}~\bibnamefont {{Van Bremen}}}, \bibinfo {author}
		{\bibfnamefont {K.}~\bibnamefont {Sotthewes}},\ and\ \bibinfo {author}
		{\bibfnamefont {H.~J.}\ \bibnamefont {Zandvliet}},\ }\bibfield  {title}
	{\bibinfo {title} {{Singularities and topologically protected states in
				twisted bilayer graphene}},\ }\bibfield  {journal} {\bibinfo  {journal}
		{Applied Physics Letters}\ }\textbf {\bibinfo {volume} {116}},\ \href
	{https://doi.org/10.1063/1.5135071} {10.1063/1.5135071} (\bibinfo {year}
	{2020})\BibitemShut {NoStop}%
	\bibitem [{\citenamefont {Wang}\ \emph {et~al.}(2015)\citenamefont {Wang},
		\citenamefont {Gao}, \citenamefont {Wen}, \citenamefont {Han}, \citenamefont
		{Taniguchi}, \citenamefont {Watanabe}, \citenamefont {Koshino}, \citenamefont
		{Hone},\ and\ \citenamefont {Dean}}]{Wang2015}%
	\BibitemOpen
	\bibfield  {author} {\bibinfo {author} {\bibfnamefont {L.}~\bibnamefont
			{Wang}}, \bibinfo {author} {\bibfnamefont {Y.}~\bibnamefont {Gao}}, \bibinfo
		{author} {\bibfnamefont {B.}~\bibnamefont {Wen}}, \bibinfo {author}
		{\bibfnamefont {Z.}~\bibnamefont {Han}}, \bibinfo {author} {\bibfnamefont
			{T.}~\bibnamefont {Taniguchi}}, \bibinfo {author} {\bibfnamefont
			{K.}~\bibnamefont {Watanabe}}, \bibinfo {author} {\bibfnamefont
			{M.}~\bibnamefont {Koshino}}, \bibinfo {author} {\bibfnamefont
			{J.}~\bibnamefont {Hone}},\ and\ \bibinfo {author} {\bibfnamefont {C.~R.}\
			\bibnamefont {Dean}},\ }\bibfield  {title} {\bibinfo {title} {{Evidence for a
				fractional fractal quantum Hall effect in graphene superlattices}},\ }\href
	{https://doi.org/10.1126/science.aad2102} {\bibfield  {journal} {\bibinfo
			{journal} {Science}\ }\textbf {\bibinfo {volume} {350}},\ \bibinfo {pages}
		{1231} (\bibinfo {year} {2015})},\ \Eprint {https://arxiv.org/abs/1505.07180}
	{arXiv:1505.07180} \BibitemShut {NoStop}%
	\bibitem [{\citenamefont {Kim}\ \emph {et~al.}(2017)\citenamefont {Kim},
		\citenamefont {DaSilva}, \citenamefont {Huang}, \citenamefont {Fallahazad},
		\citenamefont {Larentis}, \citenamefont {Taniguchi}, \citenamefont
		{Watanabe}, \citenamefont {LeRoy}, \citenamefont {MacDonald},\ and\
		\citenamefont {Tutuc}}]{Kim2017}%
	\BibitemOpen
	\bibfield  {author} {\bibinfo {author} {\bibfnamefont {K.}~\bibnamefont
			{Kim}}, \bibinfo {author} {\bibfnamefont {A.}~\bibnamefont {DaSilva}},
		\bibinfo {author} {\bibfnamefont {S.}~\bibnamefont {Huang}}, \bibinfo
		{author} {\bibfnamefont {B.}~\bibnamefont {Fallahazad}}, \bibinfo {author}
		{\bibfnamefont {S.}~\bibnamefont {Larentis}}, \bibinfo {author}
		{\bibfnamefont {T.}~\bibnamefont {Taniguchi}}, \bibinfo {author}
		{\bibfnamefont {K.}~\bibnamefont {Watanabe}}, \bibinfo {author}
		{\bibfnamefont {B.~J.}\ \bibnamefont {LeRoy}}, \bibinfo {author}
		{\bibfnamefont {A.~H.}\ \bibnamefont {MacDonald}},\ and\ \bibinfo {author}
		{\bibfnamefont {E.}~\bibnamefont {Tutuc}},\ }\bibfield  {title} {\bibinfo
		{title} {{Tunable moir{\'{e}} bands and strong correlations in
				small-twist-angle bilayer graphene}},\ }\href
	{https://doi.org/10.1073/pnas.1620140114} {\bibfield  {journal} {\bibinfo
			{journal} {Proceedings of the National Academy of Sciences of the United
				States of America}\ }\textbf {\bibinfo {volume} {114}},\ \bibinfo {pages}
		{3364} (\bibinfo {year} {2017})}\BibitemShut {NoStop}%
	\bibitem [{\citenamefont {Kerelsky}\ \emph {et~al.}(2019)\citenamefont
		{Kerelsky}, \citenamefont {McGilly}, \citenamefont {Kennes}, \citenamefont
		{Xian}, \citenamefont {Yankowitz}, \citenamefont {Chen}, \citenamefont
		{Watanabe}, \citenamefont {Taniguchi}, \citenamefont {Hone}, \citenamefont
		{Dean}, \citenamefont {Rubio},\ and\ \citenamefont
		{Pasupathy}}]{Kerelsky2019}%
	\BibitemOpen
	\bibfield  {author} {\bibinfo {author} {\bibfnamefont {A.}~\bibnamefont
			{Kerelsky}}, \bibinfo {author} {\bibfnamefont {L.~J.}\ \bibnamefont
			{McGilly}}, \bibinfo {author} {\bibfnamefont {D.~M.}\ \bibnamefont {Kennes}},
		\bibinfo {author} {\bibfnamefont {L.}~\bibnamefont {Xian}}, \bibinfo {author}
		{\bibfnamefont {M.}~\bibnamefont {Yankowitz}}, \bibinfo {author}
		{\bibfnamefont {S.}~\bibnamefont {Chen}}, \bibinfo {author} {\bibfnamefont
			{K.}~\bibnamefont {Watanabe}}, \bibinfo {author} {\bibfnamefont
			{T.}~\bibnamefont {Taniguchi}}, \bibinfo {author} {\bibfnamefont
			{J.}~\bibnamefont {Hone}}, \bibinfo {author} {\bibfnamefont {C.}~\bibnamefont
			{Dean}}, \bibinfo {author} {\bibfnamefont {A.}~\bibnamefont {Rubio}},\ and\
		\bibinfo {author} {\bibfnamefont {A.~N.}\ \bibnamefont {Pasupathy}},\
	}\bibfield  {title} {\bibinfo {title} {{Maximized electron interactions at
				the magic angle in twisted bilayer graphene}},\ }\href
	{https://doi.org/10.1038/s41586-019-1431-9} {\bibfield  {journal} {\bibinfo
			{journal} {Nature}\ }\textbf {\bibinfo {volume} {572}},\ \bibinfo {pages}
		{95} (\bibinfo {year} {2019})}\BibitemShut {NoStop}%
	\bibitem [{\citenamefont {Brihuega}\ and\ \citenamefont
		{Yndurain}(2018)}]{Brihuega2018}%
	\BibitemOpen
	\bibfield  {author} {\bibinfo {author} {\bibfnamefont {I.}~\bibnamefont
			{Brihuega}}\ and\ \bibinfo {author} {\bibfnamefont {F.}~\bibnamefont
			{Yndurain}},\ }\bibfield  {title} {\bibinfo {title} {{Selective Hydrogen
				Adsorption in Graphene Rotated Bilayers}},\ }\href
	{https://doi.org/10.1021/acs.jpcb.7b05085} {\bibfield  {journal} {\bibinfo
			{journal} {Journal of Physical Chemistry B}\ }\textbf {\bibinfo {volume}
			{122}},\ \bibinfo {pages} {595} (\bibinfo {year} {2018})}\BibitemShut
	{NoStop}%
\end{thebibliography}

%

\end{document}